\documentclass[twoside]{article}

\usepackage{PRIMEarxiv}

\usepackage[utf8]{inputenc} % allow utf-8 input
\usepackage[T1]{fontenc}    % use 8-bit T1 fonts
\usepackage{hyperref}       % hyperlinks
\usepackage{url}            % simple URL typesetting
\usepackage{booktabs}       % professional-quality tables
\usepackage{amsfonts}       % blackboard math symbols
\usepackage{nicefrac}       % compact symbols for 1/2, etc.
\usepackage{microtype}      % microtypography
\usepackage{lipsum}
\usepackage{fancyhdr}       % header
\usepackage{graphicx}       % graphics
\graphicspath{{media/}}     % organize your images and other figures under media/ folder
\usepackage[table]{xcolor}
\usepackage{multirow}
\usepackage{makecell}

\title{User Modeling in Model-Driven Engineering: A Systematic Literature Review\thanks{This project is supported by the Luxembourg National Research Fund (FNR) PEARL program, grant agreement 16544475}}

\author{
  Aaron Conrardy\\
  Luxembourg Institute of Science and Technology\\ University of Luxembourg\\
  Luxembourg\\
  aaron.conrardy@list.lu \\
\And
  Alfredo Capozucca\\
  University of Luxembourg\\
  Luxembourg\\
  alfredo.capozucca@uni.lu 
\And
  Jordi Cabot\\
  Luxembourg Institute of Science and Technology\\ University of Luxembourg\\
  Luxembourg\\
  jordi.cabot@list.lu 
}

\usepackage{graphicx} %

\usepackage{hyperref}
\begin{document}
\newpage
\thispagestyle{empty}
\vspace*{\fill}
\begin{center}
    {\Huge \textbf{This is a non-reviewed preprint. For the published version, which is available as open access, visit \url{https://doi.org/10.5381/jot.2025.24.2.a12}}}
\end{center}
\vspace*{\fill}
\newpage

\maketitle

\begin{abstract}
    In software applications, user models can be used to specify the profile of the typical users of the application, including personality traits, preferences, skills, etc. In theory, this would enable an adaptive application behavior that could lead to a better user experience.
Nevertheless, user models do not seem to be part of standard modeling languages nor common in current model-driven engineering (MDE) approaches. 
In this paper, we conduct a systematic literature review to analyze existing proposals for user modeling in MDE and identify their limitations. The results showcase that there is a lack of a unified and complete user modeling perspective. Instead, we observe a lot of fragmented and partial proposals considering only simple user dimensions and with lack of proper tool support. This limits the implementation of richer user interfaces able to better support the user-specific needs. 
Therefore, we hope this analysis triggers a discussion on the importance of user models and their inclusion in MDE pipelines.
Especially in a context where, thanks to the rise of AI techniques, personalization, based on a rich number of user dimensions, is becoming more and more of a possibility.
\end{abstract}

%\textcolor{red}{TODO; Maybe make about the difference between models that propose a generic property attribute and ones that are more specific. While generic attributes are nice, there is the need to edit the other part. In a optimal world, we could have both, we propose attributes that are predefined and can be set. But let a generic property open. Anyone is free to add them. But could add them to the original repository and then the added property is forever part of the common model.} \\

\section{Introduction}\label{section1}
Model-Driven Engineering (MDE) (and its variations such as model-based software engineering, low-code software development, etc.) focuses on the usage of modeling for the development of complex software applications, with the purpose of increasing efficiency and effectiveness in software development \cite{bookmodeling}. %MDE enables the complete or partial modeling of complex applications and provides higher abstraction levels, facilitating a better understanding of the underlying components but also promoting re-usability. 

In any of these complex applications, the human aspects play a key role \cite{humanspects}. 
Indeed, if one talks about modeling a complete software application, modeling the user(s) of the application is just as important. Generally, applications aim to help the users achieve their goals while providing them with the best possible user experience. For that purpose, both the users' behaviors but also their characteristics (preferences, cultural background, accessibility, etc.) need to be modeled, particularly the latter as reflected by the need for equality in today's software world \cite{inequality}.

Additionally, emerging technologies, both hardware (e.g. IoT devices, VR/AR or generally more powerful components) and software (e.g. machine learning (ML) or large language models (LLMs)), allow for the creation of new types of applications, with a higher degree of personalization, which was not the case in the past \cite{aijordi}. 
For example, a recent study has attempted to infer the fatigue level of the user by combining the data from a smartwatch and ML models \cite{smartwatch}, showing that new ways of profiling and adapting to users are emerging.
Indeed, we see richer types of interactions and new possibilities for adaptation (e.g.  tuning the response text of an AI-assistant to the  personality or language skills of the user asking the question) in these new modalities of human-computer interaction.

Yet, there appears to be a lack of focus on modeling users in existing MDE approaches \cite{humanspects, liebel_human_2024, userexperiencefuture, humanfactors}, which hampers leveraging these new possibilities in MDE-driven development processes. 
To evaluate the state of user modeling in MDE, we performed a systematic literature review (SLR), collecting different attempts and methods to model users, the applications they were applied in and the provided tool support. Based on these results, we argue for the need of a renewed emphasis on user modeling and, in particular, the need for a unified user model, covering a rich number of dimensions, to facilitate the specification of complex user profiles. These models would then enable code-generation processes targeting the development of applications with adaptive features such as intelligent user interfaces. 

%, and The results help to find a way towards a unified user model and showcase the need to include newer technologies when modeling end-users in MDE.

The rest of the paper is structured as follows: in Section \ref{background}, we go over relevant definitions. Section \ref{motivation} defines the need for an SLR. Section \ref{methodology} describes the process of the SLR, providing the research questions, used query and libraries, inclusion/exclusion criteria, data extraction and data synthesis. The results are presented in Section \ref{results} and discussed in Section \ref{discussion}. Section \ref{validity} describes the threats to validity of the performed review. Conclusions and next steps about how to use the SLR's results are given in Section \ref{conclusion}.

\section{Preliminary definitions}\label{background}

A \textit{user model} describes one or more application users, also called end-users, including information on the user dimensions that are relevant to the application at hand \cite{RICH1979329, purificato2024usermodelinguserprofiling}. Specifically, dimensions intrinsic to the users, such as preferences, goals, behaviors or demographic details, are represented in a structured manner as part of a user model. In theory, these models should then let the application know its user(s) and provide services adapted to such user dimensions. 

In MDE, this definition remains largely valid \cite{userexperiencefuture}. 
Only the importance of a concrete formalization increases. 
Indeed, a core element in MDE is the specification of models using a specific modeling language (or domain-specific language) formalized via a metamodel or a grammar. Both define the set of elements that could be part of a user model and how those elements can be combined among them.

The term \textit{user modeling} refers to the creation process of a user model in an application \cite{purificato2024usermodelinguserprofiling}. 
Essentially, given a user and the different dimensions of the user (meta)model, user modeling involves assigning to a user a set of values for each dimension. This is also known as the \textit{user profile}.
One usually differentiates between static user dimensions that remain the same during a given runtime session (e.g. age), and dynamic ones, that could continuously be updated (e.g. mood).

\section{Motivation}\label{motivation}

Proposing metamodels or grammars for different modeling perspectives is recurrent in the MDE community \cite{metamodelcontribution}.
While a call to establish a unified user model has been made in the past \cite{userexperiencefuture}, the inclusion of human factors in MDE still seems to be an open problem \cite{liebel_human_2024, humanfactors}.

User models, user features, human factors have all been studied in past MDE research works but the domain lacks both a unified user model and methods on how incorporate it during the engineering process.

This need for a unified user model along with the MDE processes to exploit it is exacerbated by the advent of new AI techniques. Therefore, to determine how user models can be used not only to develop AI systems, but also how AI can contribute to define such user models, it is first required to know where we are standing.

This is exactly the goal of the present SLR. To the best of our knowledge, there is no systematic study that attempts to analyse the state of user models in MDE.
The most relevant work we could find was a recent study from \cite{P29}.
They performed a targeted literature review to find dimensions of user models that can leveraged for intelligent user interfaces, only taking dimensions into account that can be learned through the user interaction. This led to them restricting the scope of their search results by focusing on a specific type of application and dimensions.
Their results do not cover the other RQs targeted in this SLR, presented in the next section. 

\section{Research Methodology}\label{methodology}
This section describes the process followed to conduct  the SLR, which mainly adhered to the well-established guidelines defined by Kitchenham \cite{kitchenham2007guidelines}. SLRs have the purpose to provide an overview of the existing contributions for a specific research question or topic, while guaranteeing an objective, reproducible and exhaustive result. 
The guidelines expect establishing  first the need for the SLR, which was done in Section \ref{motivation}.
% include date when search was conducted
%mention that each time a paper does the same, not worth snowballing the same
%we see that while papers attempt to have a user model in their models, it is underspecified and very generic
%only keep metamodels that are well defined and contain the necessary information for instantiation (metamodel )

\subsection{Research Questions}
We defined four research questions (RQ) that focus on the aspects that help to grasp the state of the art of user modeling in MDE and identify gaps: 

\textbf{RQ1: Which dimensions of the user are modeled?}

RQ1 explores the user dimensions that are represented in the existing user models.
As mentioned in Section \ref{background}, these are the intrinsic dimensions of users.
We excluded other  more extrinsic dimensions (e.g. the user current location or connected devices)  which define more the context the users find themselves in more than the users' profiles.
Beyond a pure list of dimensions, RQ1 also aims to establish the popularity of such dimensions, based on the number of sources that mention each dimension. % recurrence of each dimensions, we inferred the importance and popularity attributed to them, revealing the ones that are not modeled as often.

\textbf{RQ2: How is the user model used?}

RQ2 investigates the actual usage scenarios of user models in software applications. This includes the domain of the applications (e.g., education, health, etc.) for which the user model was created and for what purpose the user model was created  (e.g., runtime adaptation, system risk analysis, etc.).

%Note: the results from the given domain of application don't seem that interesting, not many papers focus on specific domains
% should i include the bigger goal? like increased of UX? think no but lets just keep it in mind
% helps reflect per domain how relevant the human model is / which kind of human model per domain
% 

\textbf{RQ3: Are the user dimensions fixed or dynamically evolving?}

RQ3 aims to find out if current methods to populate user models are just static approaches where user values are set once at the beginning, or whether they support dynamic dimensions that are continuously adapted during the user interaction with the application. Obviously, dynamic approaches are more challenging to build but enable a more fine-grained and evolving response to the user's changing profile. 

%given an implementation or concept, how often the user's information was supposed to be profiled to set the value of a user dimension. Specifically, we distinguished between static and dynamic dimensions. The former is instantiated once and remains static during runtime, while the latter consists of models with dimensions that require to be updated at a higher rate to correctly reflect the user's state. This RQ highlighted which type of profiling is more popular and which dimensions actually require dynamic changes.
% this is not so good, i need to link this to rq1, maybe just add it to the table or create a second table with this included...

\textbf{RQ4: How are user models implemented in a given application?}

RQ4 looks at the extent of the tool support for user models. This RQ first looks at how the user model itself is formalized, i.e. are the dimensions just enumerated or formalized in some kind of grammar, (meta)model, or any other type of formalization? Then, it also assesses whether the user model can be automatically processed to generate some type of software component that would facilitate the personalization of the target applications. 

%Firstly, we analyzed how a user model was formalized in a given paper. Beyond the conceptual level, we checked the availability of implementations that process user models. 
%This not only included implementations that use the model, but also generators or transformers that support an MDE pipeline to produce components of the system. The used libraries, the corresponding produced artifacts and the availability of a modeling tool were taken into account. 
This RQ helps to determine how advanced the level of integration of the user model in current MDE pipelines, and which technologies are more often employed to support such an integration and specification.

\subsection{Search and Selection Strategy}
\begin{figure}
    \centering
    \includegraphics[width=\linewidth]{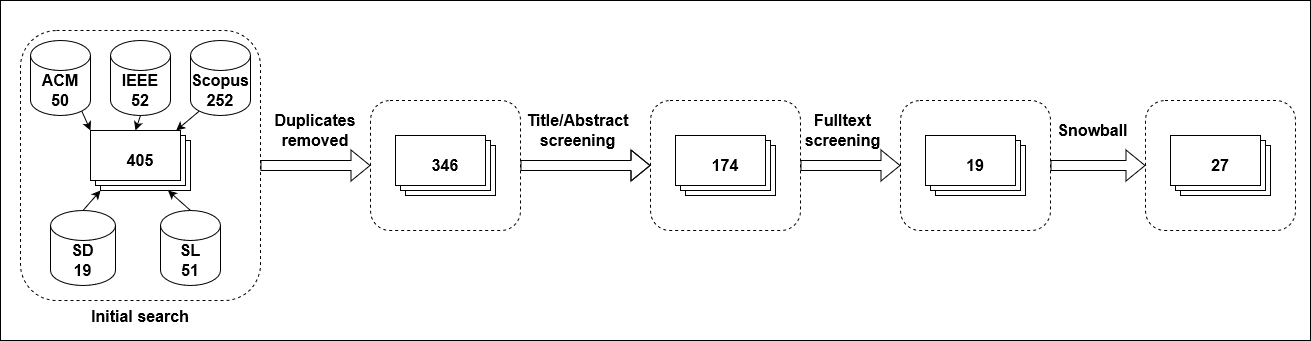}
    \caption{Search Process}
    \label{fig:search}
\end{figure}

Figure \ref{fig:search} depicts the search and selection process. 
\subsubsection{Search String}
As our goal was to find user model proposals in MDE approaches, we divided our search query into two fragments that narrow the search to the MDE field and user models respectively. The fragments consist of individual terms that were combined using the boolean operators OR and AND as follows:
\begin{center}
("model-driven" OR "model-based-software-engineering" OR "MDE" OR "MBSE" OR "MDA" OR "low-code" OR "no-code" OR "metamodel" OR "meta-model" OR "domain-specific-language" OR "DSL")\\
AND \\
    ("user/human model" OR "user/human profile" OR "user/human characteristic" OR "user/human dimension" OR "user/human attribute" OR "user/human factor" OR "user/human ontology" OR "model the end-user/user/human" OR "model end-user/user/human/person/people" OR "model(l)ing end-user/user/human/person/people") 
\end{center}

The search query has gone through numerous updates to improve the results and the number of results, the latter in an attempt to neither limit nor explode the number of results. We limited the query to look up words in the keywords, titles and abstracts. Additionally, we limit the results to journal, conferences and short papers written in English.

The MDE fragment contains the pre-fix "model-driven", the most common acronyms used in the domain, the core concepts "metamodel" and "domain-specific-language" \cite{bookmodeling} and the terms "low-code/no-code" as these are considered synonymous or variations of MDE \cite{coin, tosi2024metasciencestudyimpactlowcode, Cabot24}. We omitted the words "modeling" and "ontology" as stand-alone terms, as these led to numerous irrelevant results. 
The user model fragment consists of terms relating to the modeled characteristics ("human/user" combined with words such as "characteristics", "attributes", etc.), the produced artifacts ("human/user" and "model/ontology") and the action of modeling the users (e.g. "model the user").
%In its first version, the user model fragment consisted of the term "human" and "user" combined with any word we felt could represent the factors to be modeled, such as "characteristics", "attributes", "dimensions", etc.  
%We planned to add the terms "modeling" and "ontology" to the MDE fragment, yet found that these drastically increased the number of results. In an effort to not exclude relevant studies while still keeping a moderate amount of results, we decided to add the terms "user/human ontology" and "model(l)(ing) (the) end-user/user/human/person/people" to the human fragment.

We also took into account differences between US and UK English such as "modeling" and "modelling".

\subsubsection{Digital Libraries}\label{digitallibrary}
We decided to include the following digital libraries for conducting the search, as these are regarded as the most relevant within the software engineering domain \cite{kitchenham2007guidelines,BRERETON2007571}:
\begin{itemize}
    \item ACM Digital Library (ACM): \url{https://dl.acm.org}
    \item IEEEXplore (IEEE): \url{https://ieeexplore.ieee.org}
    \item SpringerLink (SL): \url{https://link.springer.com}
    \item ScienceDirect (SD): \url{https://www.sciencedirect.com}
    \item Scopus: \url{https://www.scopus.com}
\end{itemize}
Note that, for each library, the available search engine followed a different syntax. Thus, the defined query had to be adapted for each library. Furthermore, as neither SpringerLink nor ScienceDirect support a search based on title and abstract, we completed a full-text search. Scripts were then developed to perform the filtering at title and abstract level, which are part of the replication package for this SLR \cite{zenodo}.
The initial search was conducted on the 12.09.2024 and resulted in 346 unique (405 with duplicates) papers.
\subsubsection{Selection Strategy}
Once the initial set of unique papers was collected, we proceeded with the screening at the title/abstract level and then at the full-text level. The screening was done following a set of exclusion and inclusion criteria:
\begin{itemize}
    \item \textbf{Exclusion}:
    \begin{itemize}
        \item Not related to computer science.
        \item Not a primary study.
    \end{itemize}
    \item \textbf{Inclusion}:
    \begin{itemize}
        \item Proposes an explicit user model proposal formalized as a  metamodel, grammar, JSON schema, etc.
    \end{itemize}
\end{itemize}
During the title/abstract screening, if it was unclear whether the exclusion criteria could be applied, the paper was kept for a full-text screening. Additionally, if multiple papers from the same authors presented the same (or an updated) solution, the most recent version was kept.
In the title/abstract screening, 174 papers were kept (172 removed) and in the full-text, 19 papers were kept (155 removed).
Forward and backward snowballing \cite{snowball} was performed on the kept papers, increasing the number of total papers to 27.
For a part of the selection process, the CADIMA\footnote{\url{https://www.cadima.info/}} tool was used \cite{cadima}. Specifically, CADIMA was used for the following tasks:
\begin{enumerate}
    \item Automatically recognize duplicates from a given list of references.
    \item Propose an interface to perform the title/abstract and full-text screening.
    \item Generate an Excel file containing post-screening results.
\end{enumerate}

\subsection{Data Extraction and Synthesis}

To organize the collected data, an Excel template was created.
The filled Excel file is available in our replication package \cite{zenodo}.
For each paper, we extracted the title, the authors' names, the DOI, the venue, the year of publication and the number of citations excluding self-citations. 
Additionally, we manually added a short summary with the main takeaways. 
This summary played a key role in the next steps of the SLR as it reduced the times we had to come back to the paper for full inspection of its content.

Regarding the data gathering to answer the research questions, for RQ1, we started with a set of initial dimension categories inspired from previous works such as  \cite{Abraho2021ModelbasedIU, purificato2024usermodelinguserprofiling}. Next, for every paper, we would then extract the dimensions proposed in that paper and map them to the existing columns, adding new ones if necessary.
For RQ2, we iteratively extended the columns based on new domains or types of applications we found in the papers. 
For both RQ1 and RQ2, we took the liberty to generalize some of the extracted data, to obtain a simplified categorization.
For RQ3 and RQ4, the columns needed for the data extraction were set from the beginning and did not need any updates. 

As mentioned earlier, if the same solution appeared in multiple papers authored by the same people, the most recent one was kept. In an effort to be thorough, we still revisited all available papers related to the same solution to avoid missing any relevant information.

\section{Results}\label{results}
%\textcolor{red}{TODO; should i include additional references for works with multiple publications?} \\
 This section presents the insights gained during the analysis of the collected data.
In the end, out of the initial 346 unique papers 27 papers went through a rigorous review process. Table \ref{tab:authors} contains the list of fully reviewed papers, the first authors, their publication year and the venue they were published in. 
We observe that the venues are not limited to ones focusing on MDE or software engineering in general. Indeed, while 10 paper were published in software engineering venues (ASE, HCSE, MODELSWARD, SoSym, MEDI, MODELS, CSER, ICWE), information system venues were also a popular choice with 5 papers published (JDM, EMCIS, RCIS, UAIS). %Overall, we see an interest in MDE techniques for user modeling beyond the traditional MDE or software engineering community.

\begin{figure}[h]
    \centering
    \includegraphics[width=0.6\linewidth]{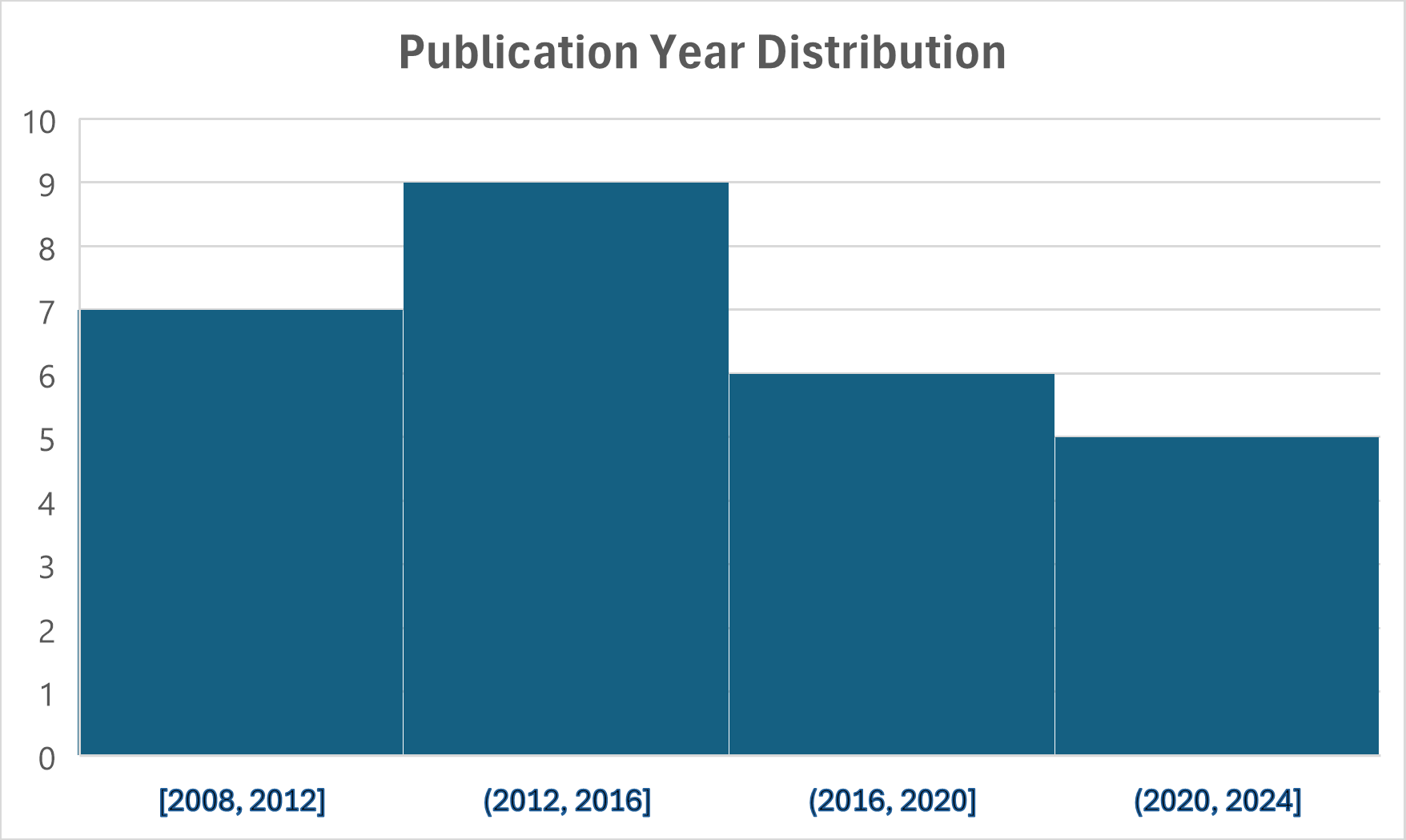}
    \caption{Histogram of publication year}
    \label{histogram}
\end{figure}

\begin{figure}[h]
    \centering
    \includegraphics[width=0.35\linewidth]{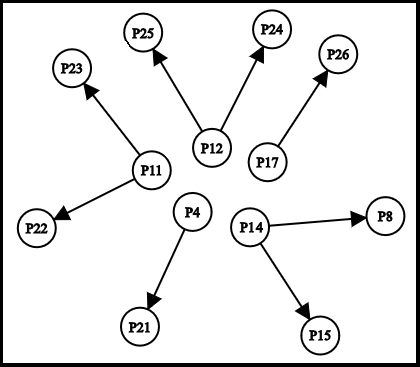}
    \caption{Paper citation graph (edge direction symbolizes citation direction)}
    \label{citation}
\end{figure}

Figure \ref{histogram} contains a histogram displaying the trend of publication over the years, where we recognize a slight decline in the number of papers. Additionally, we created a graph that showcases the citations between the selected papers in Figure \ref{citation}. Out of the 27 papers we analysed, only 8 of them appeared in the citations of 5 other papers from the chosen selection. Apparently, new proposals do not seem to deeply acknowledge and compare with previous ones at the risk of repeating themselves. 
\begin{table}
\centering
    \begin{tabular}{|c|c|c|}
        \hline
        \textbf{ID}  & \textbf{Author-Year} & \textbf{Venue}  \\
        \hline
        \hline
        P1 & \cite{P1}         & ASE\\ \hline
        P2  & \cite{P2}         & ACM SAC\\ \hline
        P3  & \cite{P3}        & ICWE \\\hline
        P4  & \cite{P4}       & IEEE ACS\\\hline
        P5  & \cite{P5}          &  WWW\\\hline
        P6  & \cite{P6}        & HCSE\\\hline
        P7  & \cite{P7}          & SysCon \\\hline
        P8  & \cite{P8}        & MDPI \\\hline
        P9  & \cite{P9}         & MODELSWARD\\\hline
        P10  & \cite{P10}         & JDM \\\hline
        P11  & \cite{P11}         & EMCIS \\\hline
        P12  & \cite{P12}         & SoSyM \\\hline
        P13  & \cite{P13}         & MEDI \\\hline
        P14  & \cite{P29}          & Multimed. Tools Appl. \\\hline
        P15  & \cite{P15}          & iiWAS \\\hline
        P16  &\cite{P16}          & MODELS\\\hline
        P17  & \cite{P17}          & Interacting with Computers\\\hline
        P18  & \cite{P18}          & ICMCS \\\hline
        P19  & \cite{P19}         & MODELS\\\hline
        P20  & \cite{P20}         & ESREL \\\hline
        P21  & \cite{P21}         & RCIS\\\hline
        P22  & \cite{P22}         & UAIS\\\hline
        P23  & \cite{P23}          & CSER \\\hline
        P24  & \cite{P25}         & SoSyM\\\hline
        P25  & \cite{P26}       & RCIS\\\hline
        P26  & \cite{P27}          & M-PREF \\\hline
        P27  &\cite{P28}        & CCC\\\hline
                
        \hline
    \end{tabular}
    \caption{Table of selected papers}
    \label{tab:authors}
    
\end{table}
% mention towards user profile ontology paper for list of attributes that are not modeled

\subsection{RQ1: Which dimensions of the user are modeled?}\label{RQ1}
%\textcolor{red}{TODO; some features such as attitudes, emotion and stress, or others might still need to be combined} \\
%\textcolor{red}{TODO; I guess it could make sense to actually explain the indiviudla features as well?, no because the readers could simply visit the papers then} \\
% in the answers for RQ3, mention that we did a categorisation ourselves, and generalized a bit. Explain then the main categories and mention that they might intersect of course
% no one models preffered personalities? 
The data extraction process provided a long list of dimensions to characterize a user. 
To better structure the results, we group the dimensions in a number of more generic categories:
\begin{itemize}
    \item \textbf{User Competencies} describe how competent a user is given a task or in general by taking into account their knowledge, learned skills and other attributes reflecting their competency.
    \item \textbf{Personality} reflect the user's personality traits, that is, stable internal characteristics of the users. 
    \item \textbf{Preferences} describe personal preferences regarding how an application does something, including the way the application interacts with the user, shows the user content, how the content is presented and more. 
    \item \textbf{Demographic Information} tackle generic information about users that are application independent and usually static such as the name, age, nationality and gender of the user. 
    \item \textbf{Accessibility} features describe accessibility needs of the user or any attribute that might affect these.
    \item \textbf{Emotions \& Mood} describe the different emotions or moods the users might experience while interacting with the application.
    \item \textbf{Goals} describe any kind of goal (related to an application or not) users might have.
    \item \textbf{Generic property} is used if in a user (meta)model, no concrete attribute is specified and a generic pair of <dimension,value> like solution is proposed.
\end{itemize}
Beyond the categories inspired from \cite{Abraho2021ModelbasedIU, purificato2024usermodelinguserprofiling}, we added the accessibility category.

Note that, dimensions mapped to one category have the potential to affect other categories or dimensions.
An example would be the possibility that the dimension "Age" could definitely affect dimensions in the "Accessibility" or "User Competencies" categories.

\begin{figure}
    \centering
    \includegraphics[width=0.6\linewidth]{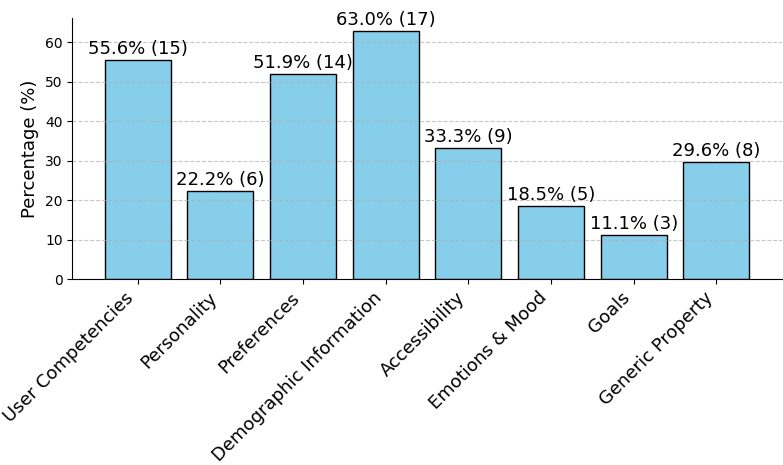}
    \caption{Coverage of dimension categories}
    \label{fig:feature-category}
\end{figure}

\begin{figure*}
    \centering
    \includegraphics[width=0.9\linewidth]{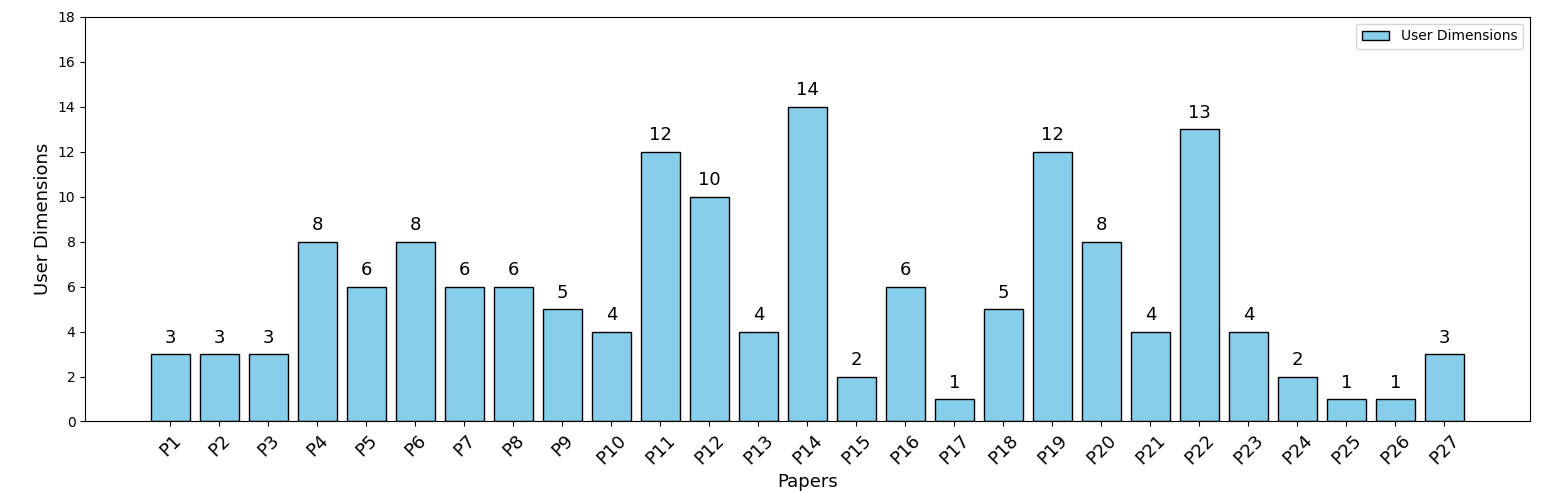}
    \caption{Number of dimensions per paper}
    \label{fig:dimensions-per-paper}
\end{figure*}

Table \ref{RQ1table} shows the list of categories and dimensions and the papers (and total number) that included such dimensions in their proposal.
A high level view of the coverage of the dimensions categories can be seen in Figure \ref{fig:feature-category} and Figure \ref{fig:dimensions-per-paper} contains the number of dimensions per paper. 

At first glance, we notice that the most popular categories of user dimensions would be the ones describing the users' competences (55.6\%), preferences (51.9\%) and demographic information (63.0\%). With a smaller yet still relevant popularity, accessibility (33.3\%) dimensions seemed to still be fairly common when modeling users. Finally, the least common dimensions were personality (22.2\%), emotions and mood (18.5\%), and goals (11.1\%). 

There were 8 papers (29.6\%) that showcased a user model that used a generic property to describe any user dimension. 
These models usually consisted of a property class that could be named and customized freely, thus the user model having the potential to include any kind of dimension. 

Excluding the generic category, the median of the number of categories covered by a specific proposal is 2, with the maximum number of categories being 5 (for papers P11, P12, P14). Regarding the dimensions themselves, the median of dimensions contained in a user model is 5, with the most being 14 (P14).

\begin{table*}

\centering

\scalebox{0.9}{
\begin{tabular}{|c|c||c|c|}
\hline
      Dimension Category & Dimension & Paper & \#Papers\\
      \hline
      \hline
      \multirow{8}{*}{User Competencies} & Role/Job & P1 P4 P5 P7 P9 P10 P12 P13 P14 P19 P23 & 11 \\ \cline{2-4}
      &Knowledge/Expertise& P1 P4 P7 P8 P11 P13 P14 P19 P23 & 9  \\ \cline{2-4}
      &Skills/Skillset& P4 P7 P8 P9 P12 P14 P16 P19 P23 & 9  \\ \cline{2-4}
      &Education&P5 P9 P14 & 3   \\ \cline{2-4}
      &Experience&P7 P9 P11 P14 P19 P20 & 6 \\ \cline{2-4}
      &Physical Capability& P20 & 1  \\ \cline{2-4}
      &Mental Capability&P19 P20  & 2  \\ \cline{2-4}
      &Reliability&P19 & 1 \\ \cline{2-4}
      
  \hline
      \multirow{5}{*}{Personality} & Attitudes & P1 P7 P11 P20 & 4 \\ \cline{2-4}
      &Emotional Resistance& P19 & 1  \\ \cline{2-4}
      &Robustness& P19 & 1   \\ \cline{2-4}
      &Motivation&P11 & 1  \\ \cline{2-4}
      &Bias&P8 & 1  \\ \cline{2-4}

        \hline
      \multirow{5}{*}{Preferences} & Generic Preference & P4 P6 P8 P10 P15 P16 P17 P27 & 8 \\ \cline{2-4}
      &Design/Presentation& P2 P18 P21 P22 & 4 \\ \cline{2-4}
      &Interaction Modality& P2 P22 & 2  \\ \cline{2-4}
      &Content& P3 P10 & 2 \\ \cline{2-4}
      &Language& P12 P18 & 2  \\ \cline{2-4}
 \hline
      \multirow{9}{*}{Demographic Information} & Name & P2 P3 P4 P5 P9 P10 P12 P13 P16 P18 P19 P21 P22 & 13\\ \cline{2-4}
      &Address& P4 P5 P14 & 3 \\ \cline{2-4}
      &Gender& P4 P5 P11 P12 P14 P16 P22 P24 & 7  \\ \cline{2-4}
      &Relationships& P3 P5 P13 P16 P19 P20 & 6 \\ \cline{2-4}
      &Nationality& P14 P18 & 2  \\ \cline{2-4}
            &Known Languages& P14 P16 P22 & 3  \\ \cline{2-4}
        &Hobbies& P14 & 1 \\ \cline{2-4}
      &Interests& P14 & 1 \\ \cline{2-4}
      &Age& P4 P11 P12 P14 P18 P27 & 6 \\ \cline{2-4}

 \hline
      \multirow{12}{*}{Accessibility} & Disability & P7 P14 P21 P23 & 4\\ \cline{2-4}
      &Body Impairements& P6 P22 & 2 \\ \cline{2-4}
      &Sight& P6 P11 P12 P22 & 4 \\ \cline{2-4}
      &Hearing& P11 P22 & 2 \\ \cline{2-4}
      &Motoric& P6 P11 & 2 \\ \cline{2-4}
      &Cognitive& P6 P22 & 2 \\ \cline{2-4}
            &Memory& P6 P20 P22 & 3 \\ \cline{2-4}
        &Attention& P6 P20 & 2\\ \cline{2-4}
      &Sensory& P6 & 1\\ \cline{2-4}
      &Speech& P22 & 1\\ \cline{2-4}
      &Mobility& P11 P22 & 2\\ \cline{2-4}
      &Physical State& P20 & 1 \\ \cline{2-4}

 \hline
      \multirow{4}{*}{Emotions \& Mood} & Emotion & P12 P14 & 2\\ \cline{2-4}
      &Mood& P12 P14 P25 &3\\ \cline{2-4}
      &Stress& P11& 1 \\ \cline{2-4}
      &Fatigue& P11 & 1 \\ \cline{2-4}
 \hline
       \multicolumn{2}{|c||}{Goals}  & P8 P14 P19  & 3  \\      
        \hline
       \multicolumn{2}{|c||}{Generic Property}  & P8 P12 P16 P21 P22 P25 P26 P27 & 8 \\ 
      \hline 
\end{tabular}

}\caption{Modeled dimensions in user model}
\label{RQ1table}
\end{table*}

\subsection{RQ2: How is the user model used?}

\begin{figure}[h]
    \centering
    \includegraphics[width=0.7\linewidth]{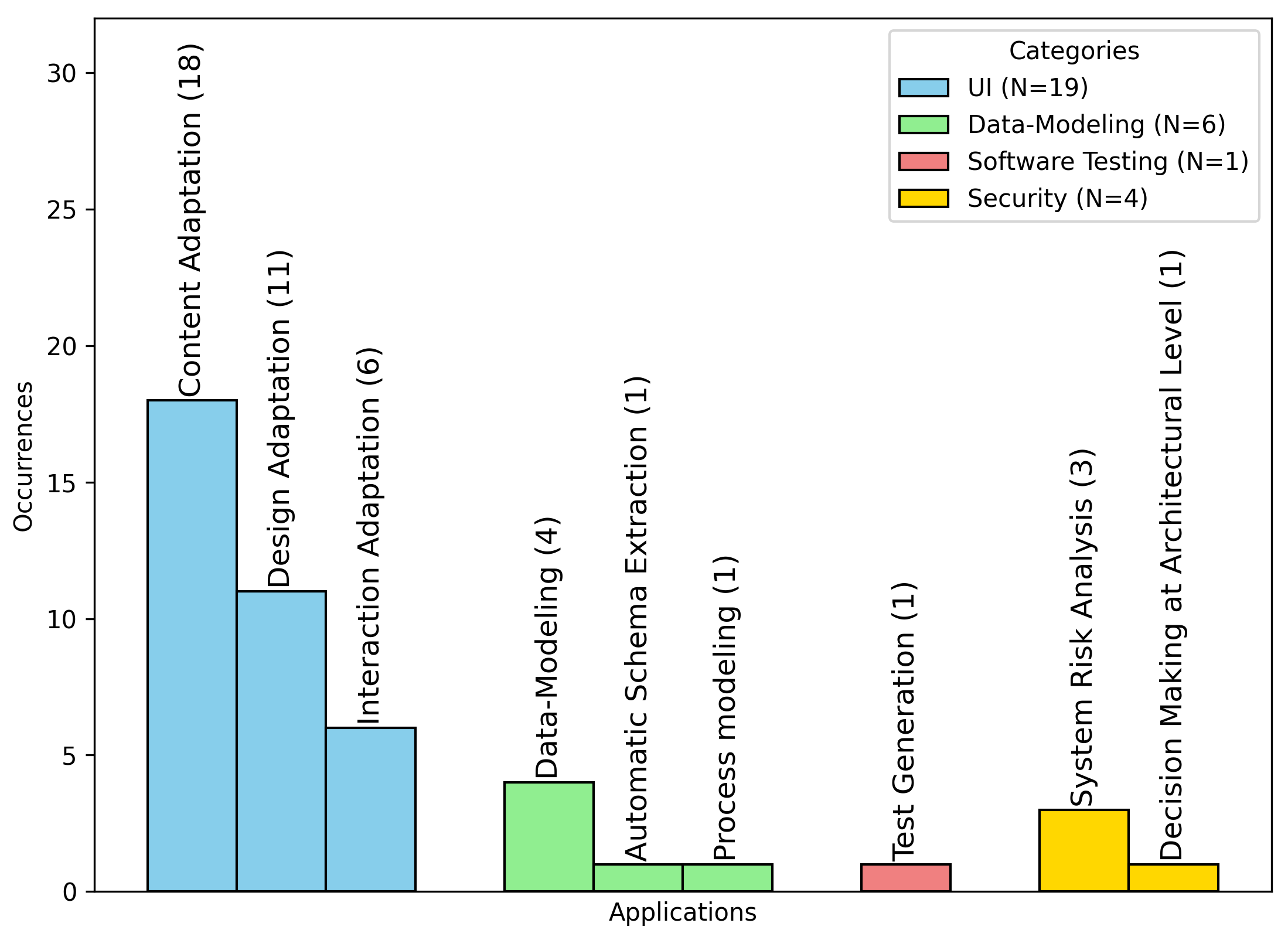}
    \caption{Distribution of application types by category}
    \label{fig:application}
\end{figure}

\begin{figure}[h]
    \centering
    \includegraphics[width=0.7\linewidth]{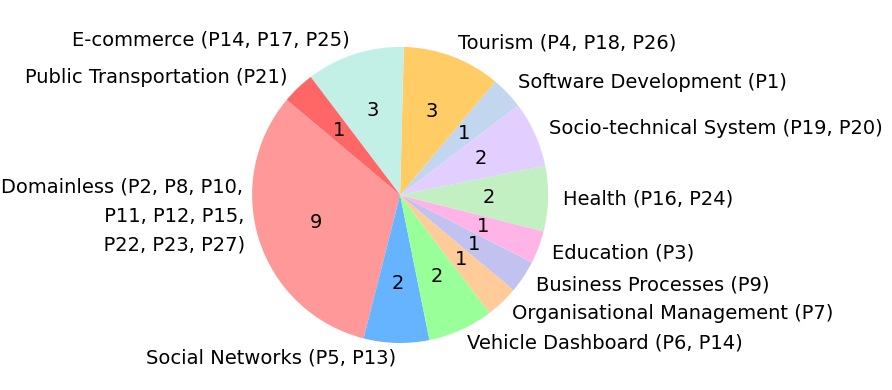}
    \caption{Distribution of domain of application}
    \label{fig:domain}
\end{figure}

Figures \ref{fig:application} and \ref{fig:domain} depict the results of the mapping of each paper proposal to a specific purpose and to a concrete domain of application respectively. 

Regarding the purpose of specifying a user model in a given work, the most popular reason was to enable adaptiveness in user interfaces (19/27). Especially the adaptation of the content itself (P2, P3, P4, P8, P9, P10, P12, P14, P15, P16, P17, P18, P21, P22, P24, P25, P26, P27) or the way that the content is presented (P2, P6, P8, P12, P14, P16, P18, P22, P24, P25, P27) were the most recurring types of adaptation. An example would be adapting which content is shown to the user, or the language of such content. Less recurrent was the adaptation of the used modality to present content to the user (P2, P12, P16, P22, P24, P25). 

Another popular purpose was modeling explicitly users to facilitate the interoperability and evolution of the system considering its users. In this context, user information is part of the data modeling efforts of the system (6/27) similar to the modeling of other system data. As an example, P7 adds a user model component to a model of organisational management systems, with the goal of providing a clearer overview of the participants in the system with a focus on the organization evolution.  

Less commonly, user models are also used to tackle security aspects of software systems (4/27) or support software testing (1/27). For the former,  some papers include the user in the model to perform risk analysis calculations to spot vulnerabilities (P19, P20, P22). P1 proposes the usage of the user model to support the development of secure software by recommending the implementation of security measures depending on the developer's competencies. The only work tackling software testing, P6, proposes the idea to combine user characteristics defined in the model with usability tests to evaluate the usability of user interfaces.

The domain of application is very diverse. In fact, a majority of the proposed applications are not tied to any specific domain (9/27).

\subsection{RQ3: Are the user characteristics fixed or dynamically evolving?} 

Table \ref{tablerq4} answers this question by listing each paper in one of the two categories.

The classification was done based on a thorough reading of the proposal where we looked for descriptions of methods on how to populate or just measure a specific user dimension, regardless if such method was implemented or not. 
%If an actual implementation of the described method was available or not did not matter to us.
Additionally, the table contains a "Unclear/Not mentioned" column that reflects cases in which in was not clear how the dimensions were to be profiled.
For example, P11 proposed various user dimensions, some we would even expect to be dynamic. Yet, no explanation is provided on how the application will profile the user, thus P11 being mapped to the "Unclear/Not mentioned" column.

As an example of each category, we can mention P1 in the static one, as P1 requires users to answer a predefined questionnaire to profile them. Once this step is completed, the dimensions of the user model are not expected to change again in the short term.
P12 would be an example of the dynamic category for its mood dimension, periodically updated by capturing pictures of the user face and updating the mood value based on the analysis of the picture.

Overall, 21 models out of the 27 propose static dimensions, with 3 of these 21 also proposing dynamic dimensions as well. No model proposed only dynamic dimensions and for 6 papers it was not clear when or how the dimensions were supposed to be profiled.

\begin{table}[h]
\centering

\scalebox{1}{
\begin{tabular}{|c||p{0.5\linewidth}|}
\hline
\textbf{Type of profiling} & \textbf{Papers} \\
\hline
\hline
\textbf{Static} & P1, P2, P3, P4, P5, P6, P7, P8, P9, P10, P12, P13, P14, P15, P17, P19, P21, P22, P24, P26, P27 \\ \hline
\textbf{Dynamic} & P3, P4, P12 \\ \hline
\textbf{Unclear/Not mentioned} & P11, P16, P18, P20, P23, P25 \\ \hline
\end{tabular}
}\caption{Type of profiling for user dimensions}\label{tablerq4}
\end{table}

\subsection{RQ4: How was the user model implemented in a given work?}
\begin{figure}
    \centering
    \includegraphics[width=0.5\linewidth]{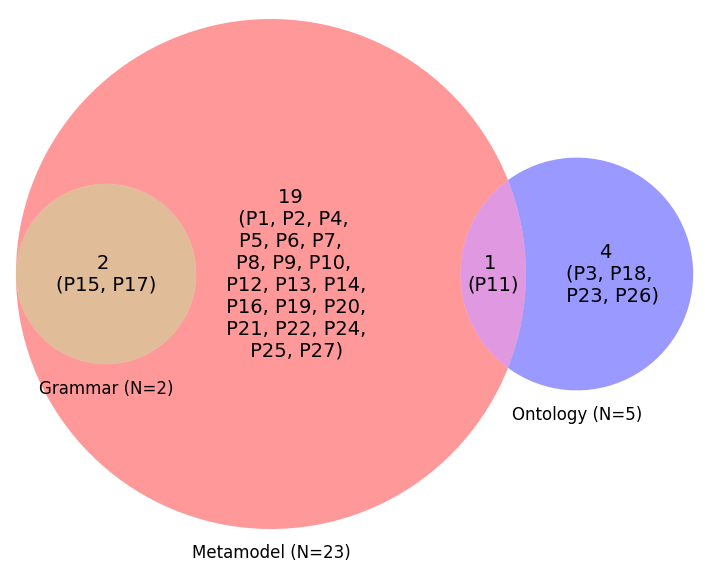}
    \caption{Venn diagram of employed formalization methods}
    \label{fig:formalization}
\end{figure}

Figure \ref{fig:formalization} summarizes the techniques to formalize the user models.
The most popular is metamodeling (23/27), typically described using UML class diagrams. 
There were 5 papers that described a user model using an ontology specified with the OWL language. 
Only 2 papers used a grammar to formalize their user model.
These 2 papers also used metamodeling to define their user model, thus no paper proposed only a grammar-based approach.

% we see a clear winner in terms of concretizing the model

Moreover, in total, 11 papers (P2, P6, P7, P12, P13, P14, P15, P19, P21, P25, P26) actually developed the applications shown in RQ2 that somehow process or leverage the user models. 
Beyond running applications, tools supporting MDE-driven development processes were proposed as well.
There are two main categories: modeling tools to enable the actual specification of users according to the user (meta)model proposed and generators able to transform these models into other software artifacts. 

\begin{figure}
    \centering
    \includegraphics[width=0.4\linewidth]{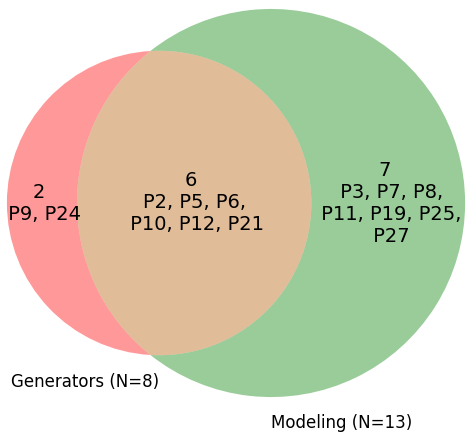}
    \caption{Available generators and modeling tools}
    \label{fig:modelingtool}
\end{figure}

This is shown in Figure \ref{fig:modelingtool}.

More specifically, the eight papers that that proposed some form of generator based on model transformation (model-to-model or model-to-text) targeted adaptive frontend code (P2, P6, P10, P12, P21, P24), the creation of user schemas to store user data (P5) or recommender systems (P9). 

The generators' transformations were implemented using technologies like ATL (4/8), Xtend (2/8), XMI (1/8) and QVT (1/8).

Regarding the modeling tools, Eclipse\footnote{\url{https://www.eclipse.org/}} is the most used as base framework (5/13), due to the easiness of creating and adding plugins to their modeling platform \cite{P12, P25}. Only 6 papers provide both a modeling tool and generators that consume the user models (P2, P5, P6, P10, P12, P21). %, which can be seen in Figure \ref{fig:modelingtool}.

In terms of numbers, 41\% of the reviewed papers propose an actual application that integrates the user model, 30\% provide generators that attempt to speed up the implementation of applications incorporating the user model and 48\% provide a modeling environment.

\section{Discussion}\label{discussion}

This section contains more in depth interpretations and reflections on the results.
\subsection{Missing dimensions in user models}\label{missing}
As showcased in Section \ref{RQ1}, the available user models already cover a wide range of user dimensions. 
Yet, we notice a trend that "simple" dimensions are modeled more often than "complex" ones. By "simple" we mean static dimensions that are easy to populate. Information such as the name, the language preference or the roles of the users tend to remain static and only require a user to enter them manually and their correctness is guaranteed, while features relating to personality, accessibility, or emotions change more frequently and are generally harder to profile. Obviously, simple dimensions are less useful when it comes to generate powerful application adaptations.

This is evident when looking at the results of RQ3, as most works describe how to work with the static data, but do not tackle the dynamic aspects even though some of the proposed dimensions could be considered dynamic (P11 models "Fatigue", yet does not mention how to profile it).  

And while current approaches propose already a good  number of dimensions, there are quite a few still missing if we look outside MDE. Indeed, user modeling is a generally a wide research topic in computer science %.  given the existence of the User Modeling conference\footnote{\url{https://www.um.org/}}, 
but also in other scientific domains such as sociology or psychology. The latter tends to focus on aspects related to the mental model or personality of humans, such as the big five models of personality \cite{big5}.
While some of proposals do try to cover personality traits, they are rather limited, covering only 1 or 2 personality related dimensions. They could be significantly enriched by looking at how users are profiled in other domains. Similarly, user models could adopt existing user taxonomies focusing on cultural attributes \cite{culture} or a wider range of emotion related attributes \cite{gumo}. 

Therefore, we can argue that user models are biased towards simple dimensions and that a stronger focus on dynamic dimensions, enriched with dimensions coming from other scientific fields, is needed to get a more complete user model. 

Some proposals try to use generic properties as way to allow for this extensibility but, in our opinion, this is a pragmatic but dangerous solution as the full application logic will depend on the interpretation of specific strings in the values of the generic property.

\begin{figure*}[t]
    \centering
    \includegraphics[width=0.9\linewidth]{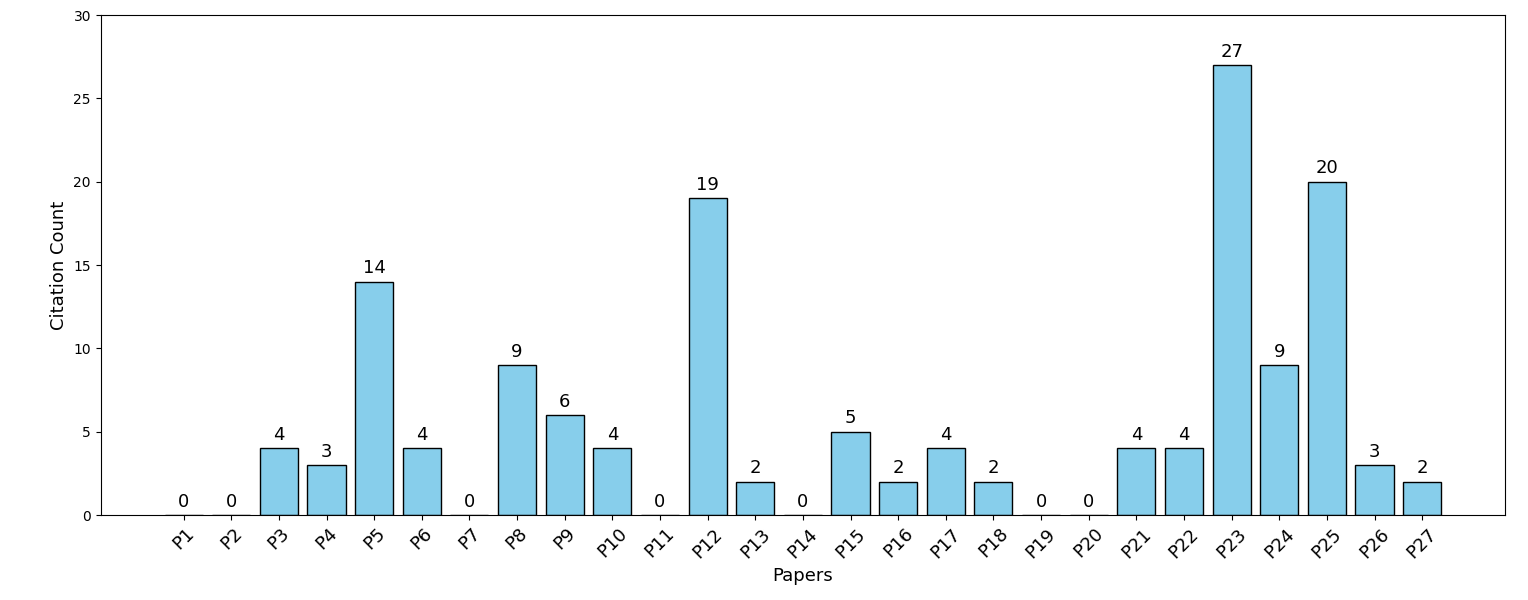}
    \caption{Citation count per paper and the amount of covered user dimensions}
    \label{fig:citationall}
\end{figure*}

\subsection{A fragmented and small community}
% in their own work
% in general because there is no interconnection or re-use 
We believe the importance of the topic doesn't really correspond to the number and impact of user modeling papers in the broad MDE community. 

In part, we believe this is because user modeling community is very fragmented, with proposals not citing and building on top of previous approaches.  See Figure \ref{citation} for an analysis of the citations among the papers in this domain. As a consequence, most papers that tackled similar topics or applications tended to create a concrete user model from scratch, thus wasting resources on re-inventing the wheel as most of the time, the final user model consists of elements present in older user models.

The lack of a unified model that could be used as core reference could also justify the small overall citation count of user modeling papers (see Figure \ref{fig:citationall}). Note that for the citation count, we removed the self-citations.

Even OMG\footnote{\url{https://www.omg.org}} standards such as UML\footnote{\url{https://www.uml.org/}} or IFML\footnote{\url{https://www.ifml.org/}}, or other standards such as UsiXML\footnote{\url{http://www.usixml.org}} do not derive or use user models from the existing literature and only provide very limited expressiveness for the definition of user models, mostly restricted to access control scenarios.  The fact that in some proposals the focus is not on the user model itself but, on the contrary, the user model is a secondary result also limits the visibility of the proposals.

We believe this situation highlights the need of a unified user model. If accepted by the main actors in this community, such user model could then be pushed as standard de facto and be adopted by other MDE approaches interested in reusing an existing user model instead of taking the time to develop their own. Potentially even making its way up to standard languages.

Otherwise, we are also giving the message that taking into account the users' needs is not a priority in the applications we build, which we believe it is not the case.

\subsection{Need for quality evaluation of user models}
While most of the analysed papers specifically mention following MDE principles and use MDE terms, only one of them (P21 that checks the consistency between the user model and the application model) performs some kind of quality evaluation of the created user models, with three more mentioning model verification as future work, without getting into details. 

Given the increasing importance of user models, we believe several of the model-based testing, verification and validation approaches could be applied to user models. For instance, we could check the intra-consistency of the model (e.g. a user that is hearing impaired is unlikely to declare audio as preferred communication modality or a user born in a certain country is likely to have some mastery of the local language(s)). At the inter-consistency level we could put in place some rules to guarantee, for instance, that user dimensions are part of the adaptations declared in the UI. On the model verification front we could make sure the user model is satisfiable, i.e. we can actually instantiate it in a way that all consistency and well-formedness rules evaluate to true (e.g. age cannot be over 200, a person cannot be in two contradictory moods at the same time, etc.).

\subsection{Interplay between machine learning and user models}
As discussed in RQ3, dynamic approaches are the exception. In part because they are more difficult to manage as they require an automatic process able to infer the values for the dynamic dimensions on a recurrent basis. 

We believe ML could be a key factor in implementing such dynamic approaches and, therefore, increasing the number of proposals focusing on dynamic dimensions (e.g. many of the approaches were created at a time where technology was not as advanced as today \cite{Abraho2021ModelbasedIU}). Technologies like Natural Language Processing (NLP) or image recognition could be used to populate several dimensions with a high degree of reliability. For instance, NLP can be used to assess the language skills of a user when interacting with the application via a chatbot, and thus, enable the adaptation of the chatbot response to a vocabulary and grammar suitable for that skill. As also seen before, image recognition could be used to infer the mood of the user.  Only 3 papers (P8, P14, P16) mention as future work to explore the usage of ML to make predictions about user preferences or classify users based on their behavior and only 1 paper (P12) actually used ML to measure the user's mood. Yet, these were more recent publications, which could indicate that the popularity and increased accessibility of ML tools will lead to an increase in using them when modeling users.

The other direction (user models as input of ML pipelines to produce AI components that are better tailored to the user profiles) is also a promising approach to increase the awareness and impact of user models. 
We have not yet seen an explicit use of user models in the ML field which is surprising given that a significant portion of machine learning tasks leverage user data \cite{purificato2024usermodelinguserprofiling} and a significant amount of MDE works already tries to get closer to the needs of machine learning \cite{NAVEED2024107423}. We believe there is an opportunity to leverage MDE to enhance ML models and algorithms with user models, by providing easy-to-use pipelines to train ML components with user data, for example, for classification purposes.

\subsection{Limited exploitation of the user models}
As revealed in RQ4, few papers come with tool support to exploit the user models. As such the return on investment for modeling user models is very low as we cannot leverage to automate other parts of the application. 

At the very least, we believe each user model proposal should come with the modeling editor but also with a number of generators or interpreters able to parse the user model and produce as output some software artefact that enhances the underlying software application. 

We are fully aware that research projects experience the valley of death phenomena  \cite{valleyofdeath} by which  artifacts developed for research projects rarely make it to the industry and rather disappear once the project ends. Indeed, investing time to create proper tool support for user models is challenging. We hope that if we first agree to try to combine better the different approaches into a unified solution we could then also joint forces to create a suite of user modeling components that could be reused and expanded later on in new projects willing to exploit user information for a variety of domains and applications. 

Eclipse could still be the basis for that, e.g. to facilitate the interoperability among the solutions, but other alternatives are possible, e.g. based on Graphical Language Server Platforms \cite{metin2023developing}.

\section{Threats to validity}\label{validity}
We will now briefly discuss aspects that may affect the validity of our results following the validity constructs defined by Wohlin et al. \cite{validity}. Note that we omit external validity, due to the nature of SLRs to possess a fixed scope, thus limiting the potential for generalization by its own nature.

\subsection{Construct validity}
Construct validity refers to the relevance of the chosen primary sources to answer the RQs.
We reduced this threat by choosing relevant libraries for the domain of software engineering, as mentioned in Section \ref{digitallibrary}. 
Additionally, the inclusion and exclusion criteria were clearly defined and discussed by all the authors.
The same goes for the search query, that contains the synonyms or terms we felt are relevant to find relevant works.
Finally, forward and backward snowballing was performed to guarantee a high coverage of results.

\subsection{Conclusion validity}
Threats to conclusion validity are concerned with how reliably we can draw conclusions about the wished or expected goal and the outcomes of the study.
In our case, the systematic process of the SLR that follows the Kitchenham guidelines already reduces this threat.
It is further reduced by the definition of the data extraction form based on the RQs.  Multiple refinements took place after joint discussions on the RQs and the data form to assure its quality and to make sure that data relevant to the RQs was extracted.

\subsection{Internal validity}
While some phases of the data extraction were developed by single reviewer, methods to ensure objective results were implemented to avoid a selection bias: 
    \begin{itemize}
        \item During the selection process, random included and excluded samples were selected by the authors to verify whether the inclusion/exclusion criteria were valid and applied correctly.
        \item A replication package is available to re-create the steps of the SLR.
         \item Regular meetings between all authors took place to discuss the validity of the following aspects of the SLR:
         \begin{itemize}
         \item The search query.
         \item The choice of digital libraries.
         \item The conclusion made based on the extracted data.
         \end{itemize}
         \item The systematic nature of the SLR following the well-established guidelines \cite{kitchenham2007guidelines} is objective in itself, such as the data extraction based on pre-defined technical questions.
    \end{itemize}

\section{Conclusion and roadmap}\label{conclusion}
In this paper, we conducted an SLR on the state of user modeling in the MDE domain. 
Results show a diverse set of disconnected proposals, covering a partial number of dimensions with an emphasis on those characteristics that are easier to profile. Moreover, most dimensions are regarded as fixed instead of allowing their dynamic evolution during the interaction with the software application.  
It is also worth noting that tool support is also rather limited, mostly limited to enabling the creation of the user models itself. 

The roadmap we hope to see in this area stems from the discussion points seen above. For instance, we believe the community should agree on a unified and re-usable user model, covering the superset of all dimensions present in the literature. Plus additional ones we could learn from user profiling in other domains (e.g. sociology).  On the technical side,  we expect to see a new generation of ML-based proposals to automatically and incrementally derive a user profile from the analysis of user interactions and a number of automatic pipelines able to transform the user information in concrete application adaptations that personalize the application to cater to the user's needs and profile.

\bibliographystyle{unsrt}  
\bibliography{references}  

\begin{thebibliography}{10}

\bibitem{bookmodeling}
Marco Brambilla, Jordi Cabot, and Manuel Wimmer.
\newblock {\em Model-Driven Software Engineering in Practice}, volume~1.
\newblock Morgan {\&} Claypool Publishers, 09 2012.

\bibitem{humanspects}
John~C. Grundy.
\newblock Impact of end user human aspects on software engineering.
\newblock In {\em Proceedings of the 16th International Conference on Evaluation of Novel Approaches to Software Engineering - ENASE}, pages 9--20. INSTICC, SciTePress, 2021.

\bibitem{inequality}
Laura Robinson, Jeremy Schulz, Grant Blank, Massimo Ragnedda, Hiroshi Ono, Bernie Hogan, Gustavo Mesch, Susan Kretchmer, Timothy Hale, Tomasz Drabowicz, Pu~Yan, Barry Wellman, Molly-Gloria Patel, Anabel Quan-Haase, Hopeton Dunn, Antonio Casilli, Paola Tubaro, Rod Carveth, and Aneka Khilnani.
\newblock Digital inequalities 2.0: Legacy inequalities in the information age.
\newblock {\em First Monday}, 07 2020.

\bibitem{aijordi}
Elena Planas, Gwendal Daniel, Marco Brambilla, and Jordi Cabot.
\newblock Towards a model-driven approach for multiexperience ai-based user interfaces.
\newblock {\em Softw. Syst. Model.}, 20(4):997–1009, August 2021.

\bibitem{smartwatch}
S.~Liu, C.~Ma, F.~Chou, M.~Cheng, C.~Wang, C.~Tsai, et~al.
\newblock Applying a smartwatch to predict work-related fatigue for emergency healthcare professionals: Machine learning method.
\newblock {\em Western Journal of Emergency Medicine: Integrating Emergency Care with Population Health}, 24(4), 2023.

\bibitem{liebel_human_2024}
Grischa Liebel, Jil Klünder, Regina Hebig, Christopher Lazik, Inês Nunes, Isabella Graßl, Jan-Philipp Steghöfer, Joeri Exelmans, Julian Oertel, Kai Marquardt, Katharina Juhnke, Kurt Schneider, Lucas Gren, Lucia Happe, Marc Herrmann, Marvin Wyrich, Matthias Tichy, Miguel Goulão, Rebekka Wohlrab, Reyhaneh Kalantari, Robert Heinrich, Sandra Greiner, Satrio~Adi Rukmono, Shalini Chakraborty, Silvia Abrahão, and Vasco Amaral.
\newblock Human factors in model-driven engineering: future research goals and initiatives for {MDE}.
\newblock {\em Software and Systems Modeling}, 23(4):801--819, August 2024.

\bibitem{userexperiencefuture}
Silvia Abrahão, Francis Bourdeleau, Betty Cheng, Sahar Kokaly, Richard Paige, Harald Stöerrle, and Jon Whittle.
\newblock User experience for model-driven engineering: Challenges and future directions.
\newblock In {\em 2017 ACM/IEEE 20th International Conference on Model Driven Engineering Languages and Systems (MODELS)}, pages 229--236, 2017.

\bibitem{humanfactors}
Judith Michael, Dominik Bork, Manuel Wimmer, and Heinrich Mayr.
\newblock Quo vadis modeling?
\newblock {\em Software and Systems Modeling}, 23:1--22, 10 2023.

\bibitem{RICH1979329}
Elaine Rich.
\newblock User modeling via stereotypes.
\newblock {\em Cognitive Science}, 3(4):329--354, 1979.

\bibitem{purificato2024usermodelinguserprofiling}
Erasmo Purificato, Ludovico Boratto, and Ernesto William~De Luca.
\newblock User modeling and user profiling: A comprehensive survey, 2024.

\bibitem{metamodelcontribution}
Richard Paige and Jordi Cabot.
\newblock What makes a good modeling research contribution?
\newblock {\em Software and Systems Modeling}, 23:1--5, 04 2024.

\bibitem{P29}
Alberto Gaspar, Miriam Gil, Ignacio Panach, and Verónica Romero.
\newblock Towards a general user model to develop intelligent user interfaces.
\newblock {\em Multimedia Tools and Applications}, 83:1--34, 01 2024.

\bibitem{kitchenham2007guidelines}
Barbara Kitchenham, Stuart Charters, et~al.
\newblock Guidelines for performing systematic literature reviews in software engineering, 2007.

\bibitem{coin}
Davide Di~Ruscio, Dimitris Kolovos, Juan Lara, Alfonso Pierantonio, Massimo Tisi, and Manuel Wimmer.
\newblock Low-code development and model-driven engineering: Two sides of the same coin?
\newblock {\em Software and Systems Modeling}, 21, 01 2022.

\bibitem{tosi2024metasciencestudyimpactlowcode}
Mauro Dalle~Lucca Tosi, Javier Luis~Cánovas Izquierdo, and Jordi Cabot.
\newblock A metascience study of the impact of low-code techniques in modeling publications, 2024.

\bibitem{Cabot24}
Jordi Cabot.
\newblock {\em The low-code handbook}.
\newblock Jordi Cabot, 2024.

\bibitem{BRERETON2007571}
Pearl Brereton, Barbara~A. Kitchenham, David Budgen, Mark Turner, and Mohamed Khalil.
\newblock Lessons from applying the systematic literature review process within the software engineering domain.
\newblock {\em Journal of Systems and Software}, 80(4):571--583, 2007.
\newblock Software Performance.

\bibitem{zenodo}
Aaron Conrardy.
\newblock User modeling in mde - slr data sheet and scripts, 2024.

\bibitem{snowball}
Claes Wohlin.
\newblock Guidelines for snowballing in systematic literature studies and a replication in software engineering.
\newblock In {\em Proceedings of the 18th International Conference on Evaluation and Assessment in Software Engineering}, EASE '14, New York, NY, USA, 2014. Association for Computing Machinery.

\bibitem{cadima}
Christian Kohl, Emma Mcintosh, Stefan Unger, Neal Haddaway, Steffen Kecke, Joachim Schiemann, and Ralf Wilhelm.
\newblock Online tools supporting the conduct and reporting of systematic reviews and systematic maps: A case study on cadima and review of existing tools.
\newblock {\em Environmental Evidence}, 7, 02 2018.

\bibitem{Abraho2021ModelbasedIU}
Silvia~Mara Abrah{\~a}o, Emilio Insfr{\'a}n, Arthur Slu{\"y}ters, and Jean Vanderdonckt.
\newblock Model-based intelligent user interface adaptation: challenges and future directions.
\newblock {\em Software and Systems Modeling}, 20:1335 -- 1349, 2021.

\bibitem{P1}
Jason Jaskolka and Brahim Hamid.
\newblock Towards the integration of human factors in collaborative decision making for secure architecture design.
\newblock In {\em Proceedings of the 37th IEEE/ACM International Conference on Automated Software Engineering}, ASE '22, New York, NY, USA, 2023. Association for Computing Machinery.

\bibitem{P2}
Chiraz El~Hog, Raoudha Ben~Djemaa, and Ikram Amous.
\newblock Profile annotation for adaptable web service description.
\newblock In {\em Proceedings of the 27th Annual ACM Symposium on Applied Computing}, SAC '12, page 1935–1940, New York, NY, USA, 2012. Association for Computing Machinery.

\bibitem{P3}
Marco Brambilla and Christina Tziviskou.
\newblock Modeling ontology-driven personalization of web contents.
\newblock In {\em 2008 Eighth International Conference on Web Engineering}, pages 247--260, 2008.

\bibitem{P4}
Zineb Aarab, Rajaa Saidi, and Moulay~Driss Rahmani.
\newblock Event-driven modeling for context-aware information systems.
\newblock In {\em 2016 IEEE/ACS 13th International Conference of Computer Systems and Applications (AICCSA)}, pages 1--8, 2016.

\bibitem{P5}
Martin Wischenbart, Stefan Mitsch, Elisabeth Kapsammer, Angelika Kusel, Birgit Pr\"{o}ll, Werner Retschitzegger, Wieland Schwinger, Johannes Sch\"{o}nb\"{o}ck, Manuel Wimmer, and Stephan Lechner.
\newblock User profile integration made easy: model-driven extraction and transformation of social network schemas.
\newblock In {\em Proceedings of the 21st International Conference on World Wide Web}, WWW '12 Companion, page 939–948, New York, NY, USA, 2012. Association for Computing Machinery.

\bibitem{P6}
Mladjan Jovanovic, Dusan Starcevic, and Zoran Jovanovic.
\newblock Bridging user context and design models to build adaptive user interfaces.
\newblock In Stefan Sauer, Cristian Bogdan, Peter Forbrig, Regina Bernhaupt, and Marco Winckler, editors, {\em Human-Centered Software Engineering}, pages 36--56, Berlin, Heidelberg, 2014. Springer Berlin Heidelberg.

\bibitem{P7}
Grace Kennedy, Farid Shirvani, William Scott, and Peter Campbell.
\newblock Managing organisational system evolution through model-based systems engineering approaches.
\newblock In {\em 2020 IEEE International Systems Conference (SysCon)}, pages 1--8, 2020.

\bibitem{P8}
Andrea Vázquez-Ingelmo, Francisco~José García-Peñalvo, Roberto Therón, and Miguel~Angel Conde.
\newblock Representing data visualization goals and tasks through meta-modeling to tailor information dashboards.
\newblock {\em Applied Sciences}, 10(7), 2020.

\bibitem{P9}
Hadjer Khider., Slimane Hammoudi., and Abdelkrim Meziane.
\newblock Business process model recommendation as a transformation process in mde: Conceptualization and first experiments.
\newblock In {\em Proceedings of the 8th International Conference on Model-Driven Engineering and Software Development - MODELSWARD}, pages 65--75. INSTICC, SciTePress, 2020.

\bibitem{P10}
Irene Garrig\'{o}s, Jes\'{u}s Pardillo, Jose-Norberto Maz\'{o}n, Jose Zubcoff, Juan Trujillo, and Rafael Romero.
\newblock A conceptual modeling personalization framework for olap.
\newblock {\em J. Database Manage.}, 23(4):1–16, October 2012.

\bibitem{P11}
Saulo Silva and Orlando Belo.
\newblock Domain and semantic modeling in the context of interactive systems development: User and device cases.
\newblock In Maria Papadaki, Paulo Rupino~da Cunha, Marinos Themistocleous, and Klitos Christodoulou, editors, {\em Information Systems}, pages 497--514, Cham, 2023. Springer Nature Switzerland.

\bibitem{P12}
Enes Yigitbas, Ivan Jovanovikj, Kai Biermeier, Stefan Sauer, and Gregor Engels.
\newblock Integrated model-driven development of self-adaptive user interfaces.
\newblock {\em Softw. Syst. Model.}, 19(5):1057–1081, September 2020.

\bibitem{P13}
Roula Karam, Piero Fraternali, Alessandro Bozzon, and Luca Galli.
\newblock Modeling end-users as contributors in human computation applications.
\newblock In Alberto Abell{\'o}, Ladjel Bellatreche, and Boualem Benatallah, editors, {\em Model and Data Engineering}, pages 3--15, Berlin, Heidelberg, 2012. Springer Berlin Heidelberg.

\bibitem{P15}
Sofiane Abbar, Mokrane Bouzeghoub, and St\'{e}phane Lopes.
\newblock Introducing contexts into personalized web applications.
\newblock In {\em Proceedings of the 12th International Conference on Information Integration and Web-Based Applications \& Services}, iiWAS '10, page 155–162, New York, NY, USA, 2010. Association for Computing Machinery.

\bibitem{P16}
Hourieh Khalajzadeh, John Grundy, and Jennifer McIntosh.
\newblock Vision: developing collaborative model-driven apps for personalised care plans.
\newblock In {\em Proceedings of the 25th International Conference on Model Driven Engineering Languages and Systems: Companion Proceedings}, MODELS '22, page 929–933, New York, NY, USA, 2022. Association for Computing Machinery.

\bibitem{P17}
Ingrid Nunes, Simone Barbosa, Donald Cowan, Simon Miles, Michael Luck, and Carlos Lucena.
\newblock Natural language-based representation of user preferences.
\newblock {\em Interacting with Computers}, 27:133--158, 01 2013.

\bibitem{P18}
Issam Elmagrouni, Mohammed Lethrech, Adil Kenzi, and Abdelaziz Kriouile.
\newblock Approach for building services-oriented systems adaptable.
\newblock In {\em 2016 5th International Conference on Multimedia Computing and Systems (ICMCS)}, pages 183--188, 2016.

\bibitem{P19}
Paul Perrotin, Salah Sadou, David Hairion, and Antoine Beugnard.
\newblock Detecting human vulnerably in socio-technical systems: a naval case study.
\newblock In {\em Proceedings of the 23rd ACM/IEEE International Conference on Model Driven Engineering Languages and Systems: Companion Proceedings}, MODELS '20, New York, NY, USA, 2020. Association for Computing Machinery.

\bibitem{P20}
Soheila~Sheikh Bahaei and Barbara Gallina.
\newblock A metamodel extension to capture post normal accidents in ar-equipped socio-technical systems.
\newblock In {\em 31st European Safety and Reliability Conference}, September 2021.

\bibitem{P21}
Ansem Ben~Cheikh, Stéphane Coulondre, Agnès Front, and Jean-Pierre Giraudin.
\newblock An engineering method for context-aware and reactive systems.
\newblock In {\em 2012 Sixth International Conference on Research Challenges in Information Science (RCIS)}, pages 1--12, 2012.

\bibitem{P22}
N.~Kaklanis, P.~Biswas, Y.~Mohamad, M.~F. Gonzalez, M.~Peissner, P.~Langdon, D.~Tzovaras, and C.~Jung.
\newblock Towards standardisation of user models for simulation and adaptation purposes.
\newblock {\em Univers. Access Inf. Soc.}, 15(1):21–48, March 2016.

\bibitem{P23}
Douglas~W. Orellana and Azad~M. Madni.
\newblock Human system integration ontology: Enhancing model based systems engineering to evaluate human-system performance.
\newblock {\em Procedia Computer Science}, 28:19--25, 2014.
\newblock 2014 Conference on Systems Engineering Research.

\bibitem{P25}
Imen Jaouadi, Raoudha Ben~Djemaa, and Han\^{e}ne Ben-Abdallah.
\newblock A model-driven development approach for context-aware systems.
\newblock {\em Softw. Syst. Model.}, 17(4):1169–1195, October 2018.

\bibitem{P26}
Vivian~Genaro Motti and Jean Vanderdonckt.
\newblock A computational framework for context-aware adaptation of user interfaces.
\newblock In {\em IEEE 7th International Conference on Research Challenges in Information Science (RCIS)}, pages 1--12, 2013.

\bibitem{P27}
Dilek Tapucu, {\"O}zg{\"u} Can, Okan Bursa, and Murat~Osman {\"U}nalir.
\newblock Metamodeling approach to preference management in the semantic web.
\newblock In {\em Multidisciplinary Workshop on Advances in Preference Handling}, pages 116--123, 2008.

\bibitem{P28}
José Bocanegra, Jaime Pavlich-Mariscal, and Angela Carrillo-Ramos.
\newblock Dmlas: A domain-specific language for designing adaptive systems.
\newblock In {\em 2015 10th Computing Colombian Conference (10CCC)}, pages 47--54, 2015.

\bibitem{big5}
Sarah~E. Babcock and Claire~A. Wilson.
\newblock {\em Big Five Model of Personality}, pages 55--60.
\newblock John Wiley \& Sons, Ltd, 2020.

\bibitem{culture}
Tom Plocher, Pei-Luen~Patrick Rau, Yee-Yin Choong, and Zhi Guo.
\newblock {\em CROSS-CULTURAL DESIGN}, chapter~10, pages 252--279.
\newblock John Wiley \& Sons, Ltd, 2021.

\bibitem{gumo}
Dominik Heckmann, Tim Schwartz, Boris Brandherm, Michael Schmitz, and Margeritta von Wilamowitz-Moellendorff.
\newblock Gumo: the general user model ontology.
\newblock In {\em Proceedings of the 10th International Conference on User Modeling}, UM'05, page 428–432, Berlin, Heidelberg, 2005. Springer-Verlag.

\bibitem{NAVEED2024107423}
Hira Naveed, Chetan Arora, Hourieh Khalajzadeh, John Grundy, and Omar Haggag.
\newblock Model driven engineering for machine learning components: A systematic literature review.
\newblock {\em Information and Software Technology}, 169:107423, 2024.

\bibitem{valleyofdeath}
Michael DiMario and Ann Hodges.
\newblock Systems engineering management in research and development valley of death.
\newblock {\em INSIGHT}, 26(3):8--14, 2023.

\bibitem{metin2023developing}
Haydar Metin and Dominik Bork.
\newblock On developing and operating glsp-based web modeling tools: Lessons learned from biguml.
\newblock In {\em 2023 ACM/IEEE 26th International Conference on Model Driven Engineering Languages and Systems (MODELS)}, pages 129--139. IEEE, 2023.

\bibitem{validity}
Claes Wohlin, Per Runeson, Martin Hst, Magnus~C. Ohlsson, Bjrn Regnell, and Anders Wessln.
\newblock {\em Experimentation in Software Engineering}.
\newblock Springer Publishing Company, Incorporated, 2012.

\end{thebibliography}

\end{document}